\documentclass[prl, twocolumn,  floatfix, superscriptaddress, aps, amsmath, amsfonts]{revtex4-1}
\usepackage[utf8]{inputenc}
\usepackage{amsmath}
\usepackage{natbib}
\usepackage{graphicx}
\usepackage{txfonts}
\usepackage{xcolor}
\usepackage{bm}

\newcommand{\add}[1]{#1}
\newcommand{\rem}[1]{}

\begin{document}
\title{Bubble dynamics for broadband microrheology of complex fluids}

\author{Brice Saint-Michel}
\affiliation{Department of Chemical Engineering, Delft University of Technology, Van der Maasweg 9, 2629HZ Delft, Netherlands}
\author{Valeria Garbin}
\email[Corresponding Author:]{v.garbin@tudelft.nl}
\affiliation{Department of Chemical Engineering, Delft University of Technology, Van der Maasweg 9, 2629HZ Delft, Netherlands}

\begin{abstract}
Bubbles in complex fluids are often desirable, and sometimes simply inevitable, in the processing of formulated products. Bubbles can rise by buoyancy, grow or dissolve by mass transfer, and readily respond to changes in pressure, thereby applying a deformation to the surrounding complex fluid. The deformation field around a stationary, spherical bubble undergoing a change in radius is simple and localised, thus making it suitable for rheological measurements. This article reviews emerging approaches to extract information on the rheology of complex fluids by analysing bubble dynamics. The focus is on three phenomena: changes in radius by mass transfer, harmonic oscillations driven by an acoustic wave, and bubble collapse. These phenomena cover a broad range of deformation frequencies, from $10^{-4}$ to $10^6$~Hz, thus paving the way to broadband microrheology using bubbles as active probes. The outstanding challenges that need to be overcome to achieve a robust technique are also discussed.
\end{abstract}

\maketitle

\section{Introduction}
Bubbles are ubiquitous in the processing and structuring of complex fluids \cite{Everitt2003, Pagani2012}. They may be undesirable and lead to poor product performance, for instance contamination of personal care products \cite{Lin1970}, or they may be key to imparting unique properties to advanced materials \cite{Fujii2017}. Bubbles are extraordinarily dynamic objects \cite{Lohse2018}: they can grow or shrink by gas diffusion, they can rise due to buoyancy, they can undergo break-up and coalescence. Owing to the compressibility of their gas core, they respond to changes in pressure, and can gently oscillate in response to acoustic waves or violently collapse in response to shock waves. Because the control of their number and size distribution is crucial in the processing and structuring of complex fluids, bubble dynamics have been studied extensively with a view to improve the stability and performance of formulated products \cite{Kloek2001}.

A useful feature of bubble dynamics is that, provided that the bubble remains spherical, the deformation field in the surrounding fluid is purely extensional \cite{Macosko1994}. This feature presents a unique opportunity to gain insights into the rheological properties of the complex fluid surrounding a bubble, because the kinematics of the deformation is completely prescribed, simple and localised -- such controlled conditions are necessary to perform a rheological measurement. A measurement of the applied stress, in combination with controlled material deformation, is also required to perform a rheological measurement. Analysis of bubble dynamics provides information to indirectly obtain the stress in the surrounding medium, with some assumptions on the constitutive behaviour \cite{Dollet2019}. In addition, the deformation time scales of bubble dynamics cover a very broad range that is normally not accessible to a single rheological technique: from the slow dynamics ($\sim10^4$~s) driven by gas diffusion \cite{Epstein1950} to the extremely fast dynamics ($\sim10^{-6}$~s) during bubble collapse \cite{Brenner2002}.

The idea of using spherical bubble deformation to probe the rheology of the surrounding material is in fact a classical concept \cite{Macosko1994}, but its full potential has remained largely unmet due to technical limitations in controlling bubble formation and subsequent deformation. Advances in experimental techniques to generate, control and deform single bubbles, and image their evolution in real time \cite{Versluis2013}, as well as advances in imaging of the microstructure of complex fluids, are now bringing this new class of methods to the forefront of soft matter and rheology \cite{Barney2020}.

This article focuses on the recent emerging applications of \emph{bubble dynamics} to the rheology of complex fluids, and highlights the current gaps that still need to be addressed to make it an accessible technique for broadband microrheology of complex fluids.

\section{Controlled deformation by bubble dynamics}
\subsection{Dynamics of spherical bubble deformation}
In this paper we focus on three dynamical phenomena that can result in the spherically symmetric deformation of a stationary, isolated bubble: changes in radius by mass transfer, harmonic oscillations of the radius driven by an acoustic wave, and bubble collapse. We do not discuss the the technique called ``cavitation rheology'', based on bubble formation by injection at the tip of a needle embedded in a soft material, which is in a more advanced state of development and has already been widely adopted, as described in a recent review \cite{Barney2020}. One advantage of the three phenomena reviewed here is that the bubble is isolated (not attached to a needle) and therefore its spherical deformation results in a purely extensional deformation of the surrounding medium, which will be described in Section~\ref{subsec:deformation}. The deformation is also localised to a region around the probe, decaying quickly away from the surface of the bubble, thus satisfying a necessary criterion for applications in microrheology \cite{Furst2017}.

The rate of deformation of the material surrounding the bubble covers several orders of magnitude between the three phenomena considered. Firstly, changes in radius by mass transfer, that is bubble growth or dissolution by exchange of gas with the surrounding medium, can occur at rates from 1~s$^{-1}$ to $10^{-4}$~s$^{-1}$ or even slower. The analysis of the change in radius of a bubble that is dissolving or growing by mass transfer, to extract information on the rheological properties of the surrounding medium, will be described in Section~\ref{sec:bubble_dissolution}. Secondly, harmonic oscillations of a bubble can be driven by an acoustic field that is close to the resonance frequency of the bubble. In the simplest case of a Newtonian fluid, and if surface tension effects are negligible, this is given by the Minnaert frequency,
\begin{equation}
\label{eq:Minnaert}
    \omega_{\mathrm{M}} = \frac{1}{R_0}\sqrt{\frac{3\kappa p_0}{\rho}} ,
\end{equation}

where $R_0$ is the equilibrium bubble radius, $\rho$ is the fluid density, $p_0$ the hydrostatic pressure, and $\kappa$ the polytropic exponent (see Section~\ref{sec:linear_osc_dynamics_equations}). Bubbles with radius ranging from $10^{-3}$~m down to $10^{-6}$~m therefore respond to acoustic waves with frequencies from $10^3$~Hz to $10^6$~Hz. The methods currently being developed to exploit this phenomenon for rheological measurements are described in Section~\ref{sec:linear_osc_dynamics}. Finally, bubble collapse is a violent phenomenon where the bubble radius can change by roughly 10 times its equilibrium value on a time-scale of tens of microseconds. The extremely large strain rates associated with bubble collapse, up to $10^6$~s$^{-1}$, can be used to probe the high-strain rate rheology of soft materials, as will be described in Section~\ref{sec:bubble_collapse}. Diffusive transport followed by cavitation have been shown to span 9 orders of magnitude of deformation rate of the surrounding material in a single experiment~\cite{Bruning2019}. The range of frequencies accessible by these phenomena therefore paves the way to a new class of techniques for broadband, active microrheology of complex fluids.

\subsection{How a bubble deforms the surrounding fluid}
\label{subsec:deformation}

A spherical bubble at a fixed position and with time-dependent radius $R(t)$ generates a purely radial velocity field. With reference to a spherical coordinate system with origin at the centre of the bubble, where $r,\theta$, and $\phi$ are the radial, azimuthal, and polar coordinates respectively, the deformation, or strain, at time $t$ can be derived based on the continuity of the velocity field at the bubble boundary in the ideal, spherically-symmetric case:
\begin{equation}
    \label{eq:v_boundary_R}
    {\bf v} (r) = v_r(r)\,  {\mathbf{e}_r} = v_r(R) \frac{R^2}{r^2} {\mathbf{e}_r} =   \dot{R} \frac{R^2}{r^2} ~{\mathbf{e}_r}\,.
\end{equation}
where the dots denote derivatives with respect to time, $\mathbf{v}=(v_r,0,0)$ is the velocity field, and the unit vector in the radial direction is $\mathbf{e}_r=(1,0,0)$. This equation assumes that the surrounding medium is incompressible, which is true for low Mach numbers $\dot{R} / c$, where $c$ is the speed of sound in the surrounding medium, and that the composition of the bubble is homogeneous.

The non-homogeneous nature of the velocity field in the surrounding medium implies that changes in the bubble radius $\dot R$ strain the surrounding medium at a rate $\dot{\bm \epsilon} = ({\bf \nabla v} + {\bf \nabla v}^{\rm T}) / 2$, where ${\bf \nabla v}$ is the (tensor) gradient of the vector field ${\bf v}$ and $(\cdot)^{\rm T}$ represents the matrix transposition operator. The spherical symmetry of the problem considerably simplifies the expression of $\dot{\bm \epsilon}$:
\begin{equation}
    \label{eq:strainrate}
    \dot{\bm \epsilon} (r) =
    \left (
        \begin{array}{ccc} 
            \frac{\partial v_r}{\partial r}& 0            & 0 \\ 
            0                              & \frac{v_r}{r}& 0 \\ 
            0                              & 0            & \frac{v_r}{r}
            \end{array} 
    \right ) = 
    \dot{R} \frac{R^2}{r^3} \left (
                        \begin{array}{ccc} 
                            -2 & 0 & 0 \\ 
                             0 & 1 & 0 \\ 
                             0 & 0 & 1 
                        \end{array} 
                    \right )
\end{equation}
The fact that the strain rate $\dot{\bm \epsilon}$ is diagonal implies that the bubble applies a pure extension or compression of material elements in the surrounding medium. This is in contrast with simple shear flow, which combines pure extension and rotation. In the case of shear flow, the strain rate $\dot{\bm \epsilon}$ is called the shear rate, and it is used to compute the shear stress in liquids; in particular, they are proportional to each other in the case of Newtonian media.

\begin{figure*}[htb]
    \centering
    \includegraphics[width=130mm]{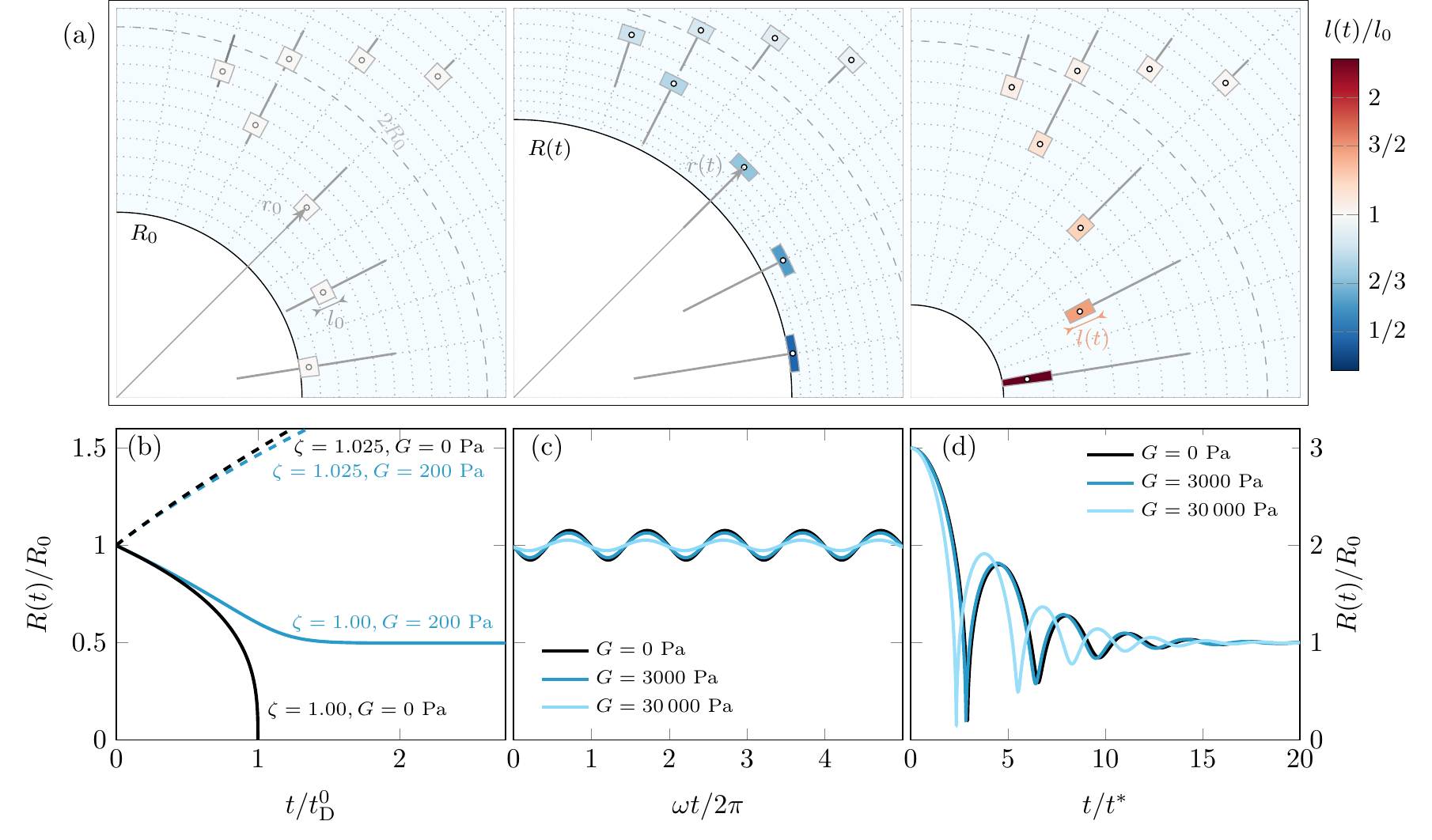}
    \caption{\textbf{Kinematics and dynamics of bubbles in fluids and soft solids.} \textbf{(a)} Kinematics of a bubble (in white) surrounded by a fluid or a soft solid (light blue) when the former is at rest (left), expands (centre) and contracts (right). The bubble surface motion pushes and pulls tracers (small circles) in the radial direction. The tracer displacement is larger close to the bubble surface (as seen on the thick grey streak lines). Material elements (boxes around the tracers) are also strained and undergo biaxial (middle) or uniaxial extension (right) when the bubble grows or shrinks. The colour of the boxes codes the stretch $l(t)/l_0$ applied to the fluid particles in the radial direction. \textbf{(b)} Time evolution of the radius of a bubble with initial radius $R_0$ due to mass transfer out of the bubble for a gas-saturated medium ($\zeta=1$) and into the bubble for a supersaturated medium ($\zeta=1.025$). \textbf{(c)} Linear oscillations of the bubble radius, $R(t)$, around the equilibrium radius $R_0$, under a small-amplitude acoustic excitation at angular frequency $\omega$. \textbf{(d)} A bubble with initial radius $3R_0$ undergoes violent collapse and subsequent rebounds, before reaching its equilibrium radius $R_0$. In (b-d) the surrounding fluid is assumed to be Newtonian when $G = 0$~Pa and otherwise combines neo-Hookean elastic and Newtonian viscous elements.}
    \label{fig:kinematics_dynamics}
\end{figure*}

In soft solids and yield-stress fluids, the stress is also a function of the total strain applied to the material elements of the surrounding medium. We can compute such a strain for an element initially at a distance $r_0$ from the centre of the bubble at rest, i.e. for $R = R_0$. This quantity is also a tensor and most continuum mechanics textbooks recommend using the Finger tensor ${\bf B}$ to quantify it~\cite{Macosko1994}:
\begin{equation}
    B_{rr} = B_{\theta\theta}^{-2} = B_{\phi\phi}^{-2} = \left (1 + \frac{R(t)^3 - R_0^3}{r_0^3} \right)^{-4/3}
\end{equation}
In the absence of deformation, the Finger tensor is not zero but rather reduces to the identity tensor ${\bf I}$. In the particular case of pure extension, we can relate the components of the Finger tensor ${\bf B}$ to the linear stretch of the fluid elements. We have in particular, for the material elements of \textbf{Figure~\ref{fig:kinematics_dynamics}(a)}, $B_{rr} = \left [ l(t) / l_0 \right ]^2$.

\textbf{Figure~\ref{fig:kinematics_dynamics}(a)} shows the properties of this strain field respectively for a bubble at rest (left panel), during its expansion (centre) and during its contraction (right panel). Both the particle displacement and the net strain ${\bf B} - {\bf I}$ are more sensitive to bubble expansion than compression and decay quickly away from the surface of the bubble.

\subsection{Bubble dynamics: relating strain and stress fields}
\label{sec:bubble_dynamics}

Both the local strain and the stress fields need to be known for bubble dynamics studies to qualify as a rheological measurement. While torque is easily measured on a rotational rheometer and readily related to the local stress in the sample, we do not directly measure the stress field around bubbles. We instead rely on the momentum balance of the whole fluid surrounding the bubble in the ${\bf e}_r$ direction, which reads~\cite{Dollet2019}:
\begin{equation}
    \label{eq:raylpless_modified}
    \rho \left (R \ddot{R} + \frac{3}{2}\dot{R}^2  \right ) = \int_{R}^\infty \left ( {\bm \nabla} \cdot {\bm \sigma} \right )_r ~{\rm d}r \,.
\end{equation}
The tensor ${\bm \sigma}$ is the Cauchy stress tensor. As it includes pressure terms, its trace is non-zero. Far away from the bubble, we assume the material is at rest, which implies that the diagonal terms of ${\bm \sigma}$ are all equal to $- p_\infty(t)$, the total (ambient and acoustic) pressure applied by the operator far away from the bubble. The spherical symmetry of the problem and the absence of torque applied to the bubble implies that $\sigma_{\theta\theta} = \sigma_{\phi\phi}$.

Equation~\ref{eq:raylpless_modified} relates an integral of the non-homogeneous stress field (right hand side) to the bubble dynamics (left hand side), yet the local stress distribution ${\bm \sigma}$ remains unknown, preventing any direct measurement of the material rheological properties. We therefore need to  choose \emph{a priori} the relation between stress and strain in the surrounding medium and try to accurately model the experimental bubble dynamics $R(t)$. For Newtonian liquids and ideal (neo-Hookean) elastic solids, these relations read:
\begin{align}
    \label{eq:constitutive_Newtonian}
    {\bm \sigma} &= -p {\bf I} + \eta \dot{\bm \epsilon} & \text{(Newtonian liquid)}\,, \\
    \label{eq:constitutive_neoHookean}
    {\bm \sigma} &= -p {\bf I} + G \left ( {\bf B} - {\bf I} \right ) & \text{(neo-Hookean solid)}\,,
\end{align}
where $p$ is a pressure term that depends on $r$ and matches the classical definition of pressure for arbitrary strains in Newtonian liquids, and only for small strains in neo-Hookean solids. The classical Newtonian viscosity $\eta$ and linear elastic modulus $G$ are considered to be free parameters of the models that are fitted to the experimental data $R(t)$. For more complex surrounding materials, such as visco-elastic liquids or solids, the stress and strain fields are related using more complex constitutive equations relating ${\bm \sigma}$, $\dot {\bm \epsilon}$ and ${\bf B}$ and additional fitting parameters.

In the specific case of Newtonian liquids and purely elastic solids, all the physical quantities in Equation~\ref{eq:raylpless_modified} are known or prescribed except for the pressure field $p(r)$. Knowing the whole pressure field is however not needed, as the right hand side term of Equation~\ref{eq:raylpless_modified} can be rewritten in the following way:
\begin{equation}
    \label{eq:divsigma_explicit}
    \int_R^{\infty} \left ( {\bm \nabla} \cdot {\bm \sigma} \right )_r ~{\rm d}r = -p_\infty(t) - \sigma_{rr}(R) + 2 \int_{R}^\infty \frac{\sigma_{rr} - \sigma_{\theta\theta}}{r} ~{\rm d}r\,. 
\end{equation}
In Equation~\ref{eq:divsigma_explicit}, the pressure contribution cancels in $\sigma_{rr} - \sigma_{\theta\theta}$ and is hence only present in the radial stress at the bubble boundary $\sigma_{rr}(R)$. This quantity can be independently solved by considering the stress balance at the bubble interface, which reads:
\begin{equation}
    \label{eq:p_bubble}
    \sigma_{rr}(R) = - p_{\rm g} + \frac{2 \gamma}{R}\,,
\end{equation}
$\gamma$ being the surface tension between the gas inside the bubble and the surrounding material, and $p_{\rm g}$ being the gas pressure in the bubble. The final step in solving Equation~\ref{eq:raylpless_modified} then consists in choosing an equation of state for the gas pressure as a function of its volume $4\pi R(t)^3 /3$, its temperature and its composition. Bubbles injected using a syringe and containing air or other non-condensible gases are usually treated as ideal gases with a homogeneous composition and are fairly simple to model. In contrast, bubbles formed through laser vaporisation of the medium are modelled as a mixture of vapour that quickly condenses at the bubble surface and an inert gas, which leads to composition gradients between the centre and the bubble edge, which necessitates additional modelling steps~\cite{Estrada2018}.

\section{Bubble dissolution or growth by mass transfer}
\label{sec:bubble_dissolution}

The dissolution or growth of gas bubbles due to mass transfer of gas from the surrounding medium [\textbf{Figure~\ref{fig:kinematics_dynamics}(b)}] is a relatively slow process for which Equation~\ref{eq:raylpless_modified} reduces to the classical isostatic criterion, ${\bm \nabla} \cdot {\bm \sigma} = {\bm 0}$.
The saturation concentration of the dissolved gas in the liquid phase, $c_{\rm s}$, is proportional to its partial pressure in the bubble following Henry's law:
\begin{equation}
    c_{\rm s} = k_{\rm H} p_{\rm g}\,.
\end{equation}
Henry's constant, $k_{\rm H}$, varies greatly between different gas/liquid pairs and depends on the temperature $T$. If we suppose that the liquid has been left long enough in contact with the same gas before conducting the experiments, we can suppose an initial saturation condition, i.e. $c(r ,t = 0) = c_{\rm s}^0 = k_{\rm H} p_0$, where $p_0$ is the ambient pressure. For bubbles small enough to be strongly affected by surface tension effects, the pressure in the gas bubble $p_g$ is greater than $p_0$, and the saturation concentration $c_{\rm s}$ exceeds $c_{\rm s}^0$ at the bubble interface, driving bubble dissolution.

For a sufficiently slow dissolution rate, the additional dissolved gas simply diffuses into the surrounding medium with a diffusion coefficient $D$. We then need to write the mass balance of the bubble and provide boundary conditions at the bubble surface and at the outer edge of the surrounding medium to derive the time evolution of the bubble radius. We choose to work with an interface that is saturated with the solute based on the gas pressure in the bubble, i.e. $c(R,t) = k_{\rm H} p_{\rm g}$. This time, the infinite surrounding medium can be initially under-saturated or super-saturated with the dissolved gas, meaning that $c(r \to \infty, t) =  c(r, t = 0) = \zeta c_{\rm s}^0$ with $\zeta$ a dimensionless constant being smaller (respectively greater) than $1$ for under- (super-) saturation conditions. For a neo-Hookean elastic surrounding medium of linear modulus $G$, the time evolution of the bubble radius then satisfies~\cite{Kloek2001}\footnote{The sign of the elastic component on the numerator has been corrected from Ref.~\citenum{Kloek2001}.}:
\begin{widetext}
\begin{equation}
    \label{eq:dissolution_Kloek}
    \frac{R_0^2}{D k_{\rm H} \mathcal{R} T } \frac{{\rm d} \tilde{R}}{{\rm d} t} = \frac{\zeta - 1 - \delta / \tilde{R}+ \delta_G (-5/4 + 1/\tilde{R} + 1/4\tilde{R}^4) }{1 + 2 \delta /3 \tilde{R} + \delta_G (5/4 - 2/3\tilde{R} + 1/12 \tilde{R}^4)} \left ( \frac{1}{\tilde{R}} + \frac{R_0}{\sqrt{\pi D t}}  \right )\,,
\end{equation}
\end{widetext}

in which $\tilde{R} = R/R_0$, $\delta = \gamma / R_0 p_0$, $\delta_G = G/ p_0$, $T$ is the temperature and $\mathcal{R}$ is the ideal gas constant. Equation~(\ref{eq:dissolution_Kloek}) is a non-linear, ordinary differential equation that can be solved numerically. Interestingly, bubble stability against dissolution more simply depends on the sign of the numerator in Equation~(\ref{eq:dissolution_Kloek}). For an initially saturated medium ($\zeta = 1$), the steady-state condition ${\rm d} \tilde{R}/{\rm d} t = 0$ reads:
\begin{equation}
    \label{eq:steadystate_dissolution}
    \frac{1}{4 \tilde{R^4}} + \left (1 - \frac{\gamma }{R_0 G} \right ) \frac{1}{\tilde{R}} - \frac{5}{4} = 0
\end{equation}
For a finite shear modulus $G$, the left-hand side of Equation~(\ref{eq:steadystate_dissolution}) is negative for $\tilde{R}= 1$ and tends to $+\infty$ for $\tilde{R} \to 0$ due to the very large positive contribution $1/4 \tilde{R^4}$ of elastic stresses for small bubbles. Consequently there always exists an equilibrium radius $0 \leq R^* \leq R_0$ in neo-Hookean solids at which dissolution stops. For $R = R^*$, the gas pressure in the bubble $p_{\rm g}$ becomes equal to $p_0$, leading to an equal solute concentration $c_{\rm s}^0$ at the bubble interface and far away from it. Neo-Hookean solids then always arrest dissolution, as seen in \textbf{Figure~\ref{fig:kinematics_dynamics}(b)}. In contrast, stresses in a fluid medium do not diverge for $\tilde{R} \to 0$ as they rather depend on the strain rate $\dot{\bm \epsilon}$ in the surrounding material, or, equivalently, in $\dot{R}$. The gas pressure in the bubble may then always exceed $p_0$ provided that the bubble dissolution rate $\dot R/ R$ is slow enough. Fluids are then incapable of halting bubble dissolution even though very viscous fluids may slow it down \cite{Kloek2001}. 

\begin{figure}[htb]
    \centering
    \includegraphics[width=80mm]{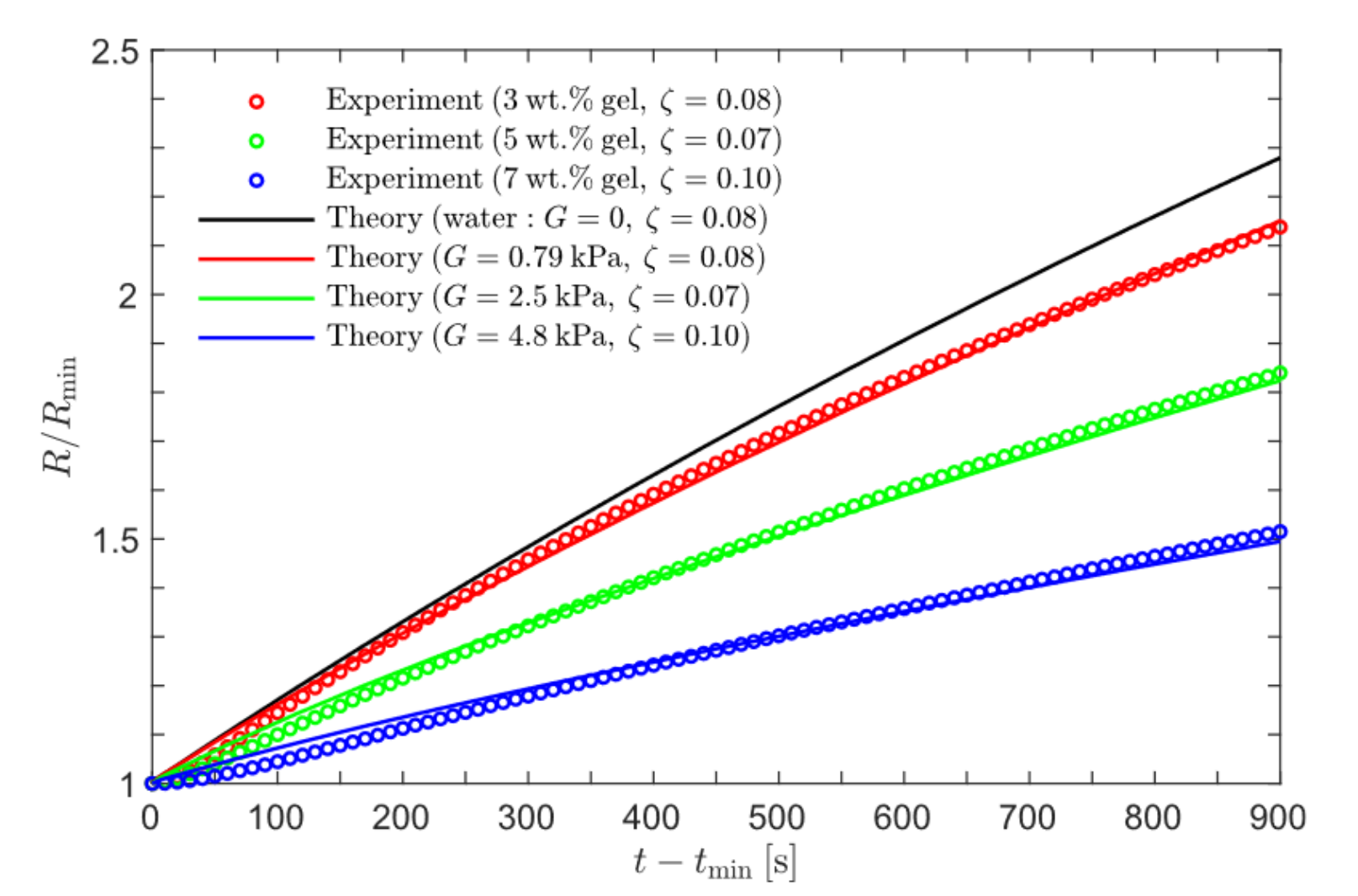}
    \caption{\textbf{Effect of medium elasticity on bubble growth by mass transfer}. A laser-generated bubble grows in an elastic medium super-saturated with air. The growth dynamics of the bubble radius $R$ is used to estimate the super-saturation coefficient $\zeta$ and the elastic modulus $G$ of the surrounding material. Reproduced with permission from Ref.~\cite{Ando2019}
    \label{fig:dissolution-growth}}
\end{figure}

When the surrounding medium is initially super-saturated with the dissolved gas, $\zeta \geq 1$, bubbles can grow by mass transfer from the surrounding medium. Elastic effects slow down this growth, and the resulting growth dynamics can be fitted to estimate both the super-saturation coefficient $\zeta$ and the elastic modulus of the medium, as shown in experiments by Ando and Shirota \cite{Ando2019}. Bubbles in a soft, viscoelastic solid were found to grow more slowly for increasing $G$ despite similar super-saturation coefficients $\zeta$, as shown in \textbf{Figure~\ref{fig:dissolution-growth}}. 

Recent experiments show that bubble dissolution can also be halted in yield-stress materials such as oleogels~\cite{Saha2020}. The situation is more complex in this case, as yield-stress materials behave as elastic solids for low applied stresses, yet flow above it. In the context of spherical bubble dynamics, the criterion for the onset of flow behaviour reads $(\sigma_{rr} - \sigma_{\theta\theta})^2 \geq 3 \sigma_{\rm Y}^2$~\cite{DeCorato2019}, $\sigma_{\rm Y}$ being a scalar quantity representing the yield strength of the material. Assuming the material behaves as an elastic solid below yielding [Equation~\ref{eq:constitutive_neoHookean}], all of it remains solid provided that 
\begin{equation}
    \label{eq:noyield_YS}
    \left [ \sigma_{rr} (R) - \sigma_{\theta\theta} (R) \right ]^2 = G^2 \left [ \left (\frac{R_0}{R} \right )^4 - \left ( \frac{R}{R_0} \right )^2 \right ]^2 < 3 \sigma_{\rm Y}^2
\end{equation}
Bubble dissolution is then arrested provided that both Equations~(\ref{eq:steadystate_dissolution}) and~(\ref{eq:noyield_YS}) are satisfied. For stiff materials for which $\gamma / R_0 G \ll 1$, Equation~(\ref{eq:steadystate_dissolution}) implies $R^* \simeq R_0$, and the no-yield criterion can be derived explicitly, leading to $\gamma / R_0 \leq \sigma_{\rm Y} / \sqrt{3}$. Materials that can withstand a typical capillary stress $\sqrt{3} \gamma / R_0$ without yielding can then halt bubble dissolution. Injecting several bubbles of different sizes and examining the smallest bubble that does not dissolve could then be used to estimate \emph{in situ} the yield stress of (gas-saturated) yield-stress materials. In practice, this picture is a bit more complicated, as most yield-stress fluids exhibit \emph{creep} behaviour for applied stresses close to $\sigma_{\rm Y}$, during which strain slowly accumulates over time. Attractive yield-stress fluids also often show thixotropic behaviour, in which the structure of the fluid reinforces --and the material constants $G$ and $\sigma_{\rm Y}$ grow-- over time at rest or under very slow flow conditions. Creep and thixotropy will therefore have antagonistic effects on the ability of yield-stress fluids to stabilise bubbles, the former promoting dissolution by allowing additional material deformation for a given applied stress, and the latter impeding dissolution when the flow is slow enough to result in an increase of $\sigma_{\rm Y}$. Surface activity of the fluid, while insufficient to stabilise bubbles by itself in Ref.~[\citenum{Saha2020}], may also play a role in slowing down or halting dissolution in conjunction with its bulk rheological properties. 

\section{Linear bubble oscillations}
\label{sec:linear_osc_dynamics}

\subsection{Bubbles as harmonic oscillators}
\label{sec:linear_osc_dynamics_equations}

The case of linear oscillations of air bubbles [Figure~\ref{fig:kinematics_dynamics}(d)] is fairly simple to model. First, on the short time scales involved and the limited oscillation amplitude, we can assume that air does not dissolve into the surrounding medium. Authors then assume~\cite{Yang2005,Gaudron2015,Hamaguchi2015,Jamburidze2017} a constant $p_g R^{3 \kappa}$ in the gas phase, choosing a polytropic exponent $1 \leq \kappa \leq 1.4$ corresponding to thermodynamic transformations that are neither fully isothermal ($\kappa = 1$) nor adiabatic ($\kappa = 1.4$ for diatomic gases). We choose to use the standard Kelvin-Voigt model for linear viscoelastic solids, that includes both a viscous and an elastic contribution. Defining $x(t) = R(t)/R_0 - 1$, we may derive the linear version of Equation~\ref{eq:raylpless_modified}~\cite{Hamaguchi2015}:
\begin{equation}
    \label{eq:raylpless_linear}
    \ddot{x} + 2 \beta \dot{x} + \omega_0^2 x = - \frac{1}{\rho R_0^2 } [p_\infty(t) - p_0]
\end{equation}
The damping term $\beta$ includes a viscous term proportional to the surrounding medium viscosity $\eta$, a thermal loss term due to thermal gradients present in the bubble, and a term accounting for acoustic damping~\cite{Prosperetti1988}. 

The natural angular frequency $\omega_0$ also depends on the surrounding medium properties,
\begin{equation}
\label{eq:natural_frequency_G}
    \omega_0^2 = \frac{3 \kappa p_0 + 2 (3 \kappa - 1 ) \gamma /R_0 + 4 G}{\rho R_0^2}\,.
\end{equation}
This frequency then deviates from the classical Minnaert resonance frequency (see Equation~\ref{eq:Minnaert}) when bubbles are sufficiently small bubbles: $R_0 \sim \gamma / p_0 \sim 1 \mu$m in water at ambient pressure, and for elastic moduli $G$ that are comparable with $p_0$, i.e. for stiff gels.

\subsection{Measuring material properties in the Fourier domain}
\label{sec:linear_bubble_spectroscopy}

In the time domain, the damping parameter $\beta$ and the natural oscillation frequency $\omega_0$ can be fitted from the free bubble oscillations, i.e. when $p_\infty(t) = p_0$ and for initial conditions $x(t = 0) \neq 0$ or $\dot{x} (t= 0) \neq 0$. An alternative approach is to study the steady-state oscillatory regime at an imposed angular frequency $\omega$. Equation~(\ref{eq:raylpless_linear}) can then be recast in the Fourier domain: 
\begin{equation}
\label{eq:resonance_curve}
     x  = \frac{1}{\rho R_0^2} \frac{p_\infty - p_0 }{\sqrt{(\omega^2 - \omega_0^2)^2 + 4 \omega^2 \beta^2}}\,,
\end{equation}

\begin{figure*}[htb]
    \centering
    \includegraphics[width=140mm]{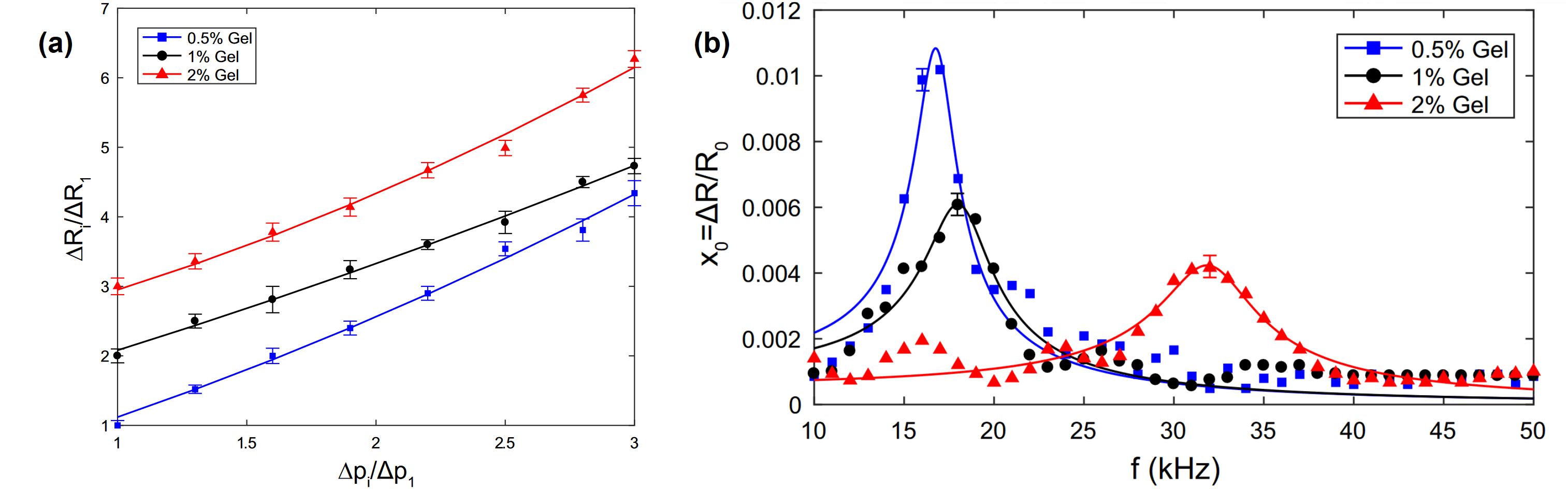}
    \caption{\textbf{Linear high-frequency rheology by acoustic bubble dynamics.} \textbf{(a)} Test to confirm the linearity of bubble oscillations and material deformation in agarose gels. The amplitude of oscillations $\Delta R$ (normalised by the smallest oscillation amplitude $\Delta R_1$) is plotted as a function of the applied pressure amplitude $\Delta p$ (normalised by the lowest applied pressure $\Delta p_1$). Symbols are experimental data and lines are quadratic fits, with black (respectively red) data series shifted upwards by $1$ (respectively 2) for clarity. \textbf{(b)} Impact of the surrounding material properties on the resonance curves. Symbols are experimental data and solid lines represent a fit from Equation~(\ref{eq:resonance_curve}). As the gel concentration is increased, the gel becomes stiffer and the resonance frequency increases [see Equation~(\ref{eq:natural_frequency_G})], and viscous damping also increases, broadening the resonance peak. Reproduced with permission from Ref.~\cite{Jamburidze2017}
    \label{fig:linear_oscillation_rheology}}
\end{figure*}

Equation~\ref{eq:resonance_curve} gives the resonance curve of an oscillating bubble in the linear regime, where it can be described as a linear harmonic oscillator \cite{Dollet2019}. It can be measured experimentally by sweeping the forcing frequency around the natural frequency $\omega_0$, and recording the (maximum) oscillation amplitude $x$ for each value of the frequency. This approach was first demonstrated by Strybulevych \emph{et al.} by measuring the acoustic response of millimetre-sized bubbles in an agar gel as a function of frequency; the approach was termed ``acoustic microrheology'' \cite{Strybulevych2009}. In a similar approach termed ``microbubble spectroscopy'', resonance curves can be recorded by directly imaging with a high-speed camera the bubble dynamics during acoustic forcing at different frequencies \cite{vanderMeer2007}. Jamburidze \emph{et al.} \cite{Jamburidze2017} also used direct imaging in agarose gels and at moderate pressure amplitudes ($p_\infty \leq 6$~kPa), to ensure that the deformation of the material remained in the linear regime (see Figure~\ref{fig:linear_oscillation_rheology}a). They extracted the viscoelastic properties of the gels from the resonance curves (see Figure~\ref{fig:linear_oscillation_rheology}b). Alternatively, Hamaguchi and Ando constructed a resonance curve by keeping the ultrasound frequency constant at 28~kHz, letting the bubble radius increase slowly due to gas transfer into the bubble \cite{Hamaguchi2015} so as to perform a ``radius sweep''. The radius sweep relies on the inverse proportionality between equilibrium radius $R_0$ and natural frequency $\omega_0$ when the magnitude of the term $2 (3 \kappa - 1 ) \gamma /R_0$ in Eq.~\ref{eq:natural_frequency_G} is sufficiently small, i.e. for sufficiently large bubbles. Resonance curves measured from radius sweeps do not rely on a prior calibration of the transducer response as a function of the frequency, making them particularly attractive in experiments where the pressure field cannot be measured independently. 

In some experiments~\cite{Jamburidze2017}, the viscous contribution to damping $\beta$ is dominant, and fitting the resonance curve directly provides an estimate of the material parameters of the constitutive model, in our case $\eta$ and $G$. In the general case~\cite{Hamaguchi2015,Saintmichel2020}, acoustic and thermal contributions have to be included in the total damping $\beta(\omega, R_0)$, which is not a constant in either frequency sweep or radius sweep experiments. In practice, reliable estimates of the viscous and thermal contributions have been determined by Prosperetti~\cite{Prosperetti1988}. Resonance curves are then fitted using the total damping $\beta(\omega,R_0)$, prescribing the viscous and thermal terms and letting the medium viscosity $\eta$ as the adjustable parameter of the fit. 

\section{Bubble collapse}
\label{sec:bubble_collapse}

\begin{figure}[htb]
    \centering
    \includegraphics[width=80mm]{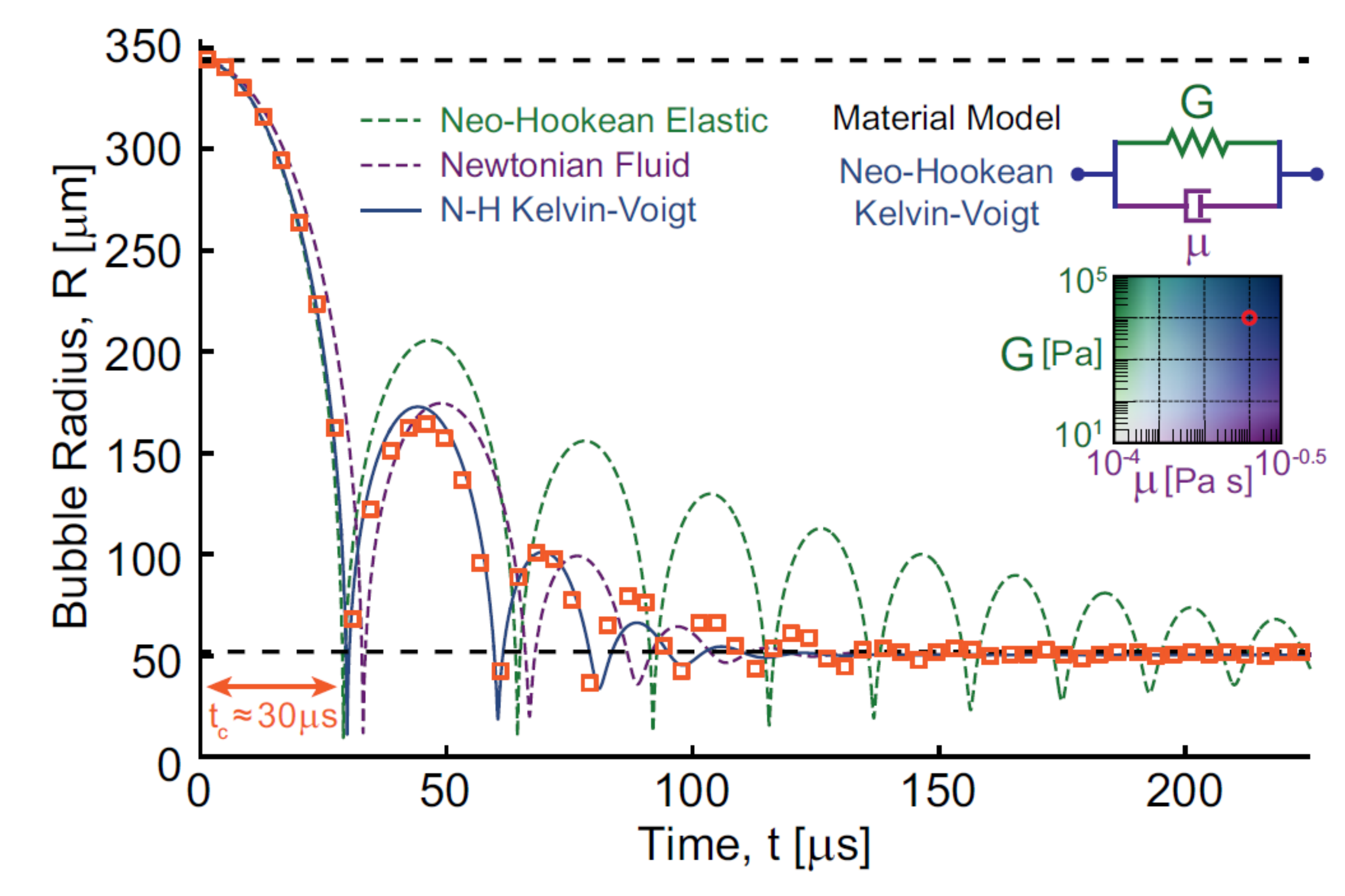}
    \caption{
    {\bf Collapse dynamics of a laser-generated bubble in a polyacrylamide gel}. The experimental data $R(t)$ (orange squares) are compared to the best fits assuming the gel behaves as a neo-Hookean elastic solid (green dashed line), a Newtonian fluid (purple dashed line) and a neo-Hookean Kelvin-Voigt --viscoelastic solid-- model (blue solid line; also see top-right schematic diagram). The fit to the viscoelastic model provides estimates of the its linear elastic modulus $G$ and viscosity $\eta$ (see right inset). Reproduced with permission from Ref.~\cite{Estrada2018}}
    \label{fig:collapse}
\end{figure}

Bubble collapse combines the high-frequency description of bubble oscillations of Section~\ref{sec:linear_osc_dynamics} with the large deformation framework derived in Section~\ref{sec:bubble_dissolution}. Bubble collapse can be achieved for instance by applying a step change in the external pressure $p_\infty$ (Rayleigh collapse), or by creating bubbles using a laser pulse to vaporise the surrounding medium, and letting them relax (Flynn collapse). Regardless of the preparation protocol and the medium properties, the complex relaxation dynamics of the bubble assumes the shape of an initial, extremely steep decrease in radius followed by a single or multiple rebounds and an eventual relaxation to an equilibrium radius (see, for instance, Figures~\ref{fig:kinematics_dynamics}d and~\ref{fig:collapse}). Conversely, the initial collapse time and the duration, number and amplitude of the rebounds depend on the type of collapse and the surrounding medium rheology, opening the way for high strain-rate characterisation of soft materials~\cite{Estrada2018,Yang2020}. Given the violent nature of the process, elasticity effects are only perceptible for stiffer materials ($G \geq 10^3$~Pa) whereas viscous effects are almost always noticeable ($\eta \geq 10^{-2}$~Pa.s)~\cite{Estrada2018}. 

Estrada \emph{et al.} have studied the collapse dynamics of a relatively stiff 10\% polyacrylamide gel and compared it with predictions from various rheological models including or combining neo-Hookean, Newtonian fluid and Maxwell fluid elements.  Figure~\ref{fig:collapse} confirms that both elastic and viscous components must be included in the model --resulting in a neo-Hookean Kelvin-Voigt material-- to accurately fit the bubble collapse dynamics. Estrada \emph{et al.}~\cite{Estrada2018} have shown that increasing the complexity of the rheological model does not significantly improve the quality of the fit to the experimental bubble collapse data. Nevertheless, Yang \emph{et al.}~\cite{Yang2020} have suggested that including additional strain-stiffening terms in the material elasticity reconciles the classical rheology measurements and the non-linear fits of bubble collapse, in agreement with the strain-stiffening behaviour generally observed in biopolymer gels~\cite{Storm2005}.

Bubble collapse is a violent, complex process and modelling efforts have been particularly cautious to evaluate all the deviations from the ideal case of adiabatic oscillations in an incompressible model fluid. Indeed, the Mach number $\dot{R}/c$ during collapse is no longer small and first-order compressibility corrections in Equation~(\ref{eq:raylpless_modified}) have to be included~\cite{Estrada2018,Yang2020}. Barajas \emph{et al.} has shown that heat transfer between the bubble and the surrounding medium must be taken into account as it noticeably affects the relaxation dynamics and the equilibrium bubble size during collapse~\cite{Barajas2017}. Lastly, Gaudron~\emph{et al.} have studied analytically and numerically the onset of non-spherical bubble oscillations during Rayleigh collapse~\cite{Gaudron2020}. Their numerical results highlight that such effects should not occur for $p_\infty \leq 15 G$. They also show that working with a finite deformation elastic framework --the neo-Hookean model-- rather than linear elasticity --Hooke's law-- strongly promotes spherical bubble oscillations, in particular for stiff materials and high applied pressures $p_\infty \sim 10^7$~Pa.

\section{Towards broadband microrheology of complex fluids using bubble dynamics}
\subsection{Current limitations of rheological measurements based on bubble dynamics}

Here we identify the main limitations that need to be overcome to achieve a robust technique. The first practical issue is the limited control offered by bubble microrheology in terms of the applied strain or stress, at variance with classical rheometers; the latter are indeed particularly efficient at applying a precise strain or stress history to the sample such as step strain, shear startup, large amplitude oscillatory shear or creep tests. During bubble dissolution, neither the strain rate or the stress at the bubble boundary are fixed, as the dynamics is rather governed by the difference in chemical potential between the bubble and the fluid far from it. Linear bubble oscillations may be viewed as small amplitude oscillatory strain experiments, yet without an explicit control on the oscillation amplitude, which depends on the inertial terms in Equation~(\ref{eq:raylpless_modified}) and the material properties. Lastly, bubble collapse experiments impose a fairly complex strain history to the material which cannot be controlled by the operator. This profile rather results from a balance between the bubble thermodynamics, fluid inertia and material rheology. 


\add{The second challenge is related to bubble formation, stability and imaging.}  Injecting or embedding spherical bubbles of a desired radius in soft materials can be challenging, especially if the material exhibits predominantly solid-like behaviour, or if it is a yield-stress fluid. Bubbles may be injected while a gel is setting \cite{Jamburidze2017} but this is not always possible. Generation of bubbles by a focused laser pulse provides the necessary spatio-temporal control on bubble formation \cite{Hamaguchi2015,Estrada2018,Ando2019}, although this is not a widely available technique. \add{Bubble dissolution will become an issue for linear oscillation experiments at $f\sim1$~MHz, corresponding to bubbles with $R_0 \sim 1\,\mu$m, as the dissolution time ($\sim 10$~ms in water) may then fall below the time needed to perform the measurements}. Furthermore it is observed that shrinking bubbles in hydrogels can leave behind a pocket of water within a solid-like gel network that remains undeformed~\cite{Hamaguchi2015}. 
\add{Fine-tuning the saturation coefficient $\zeta$ by letting samples equilibrate at a different temperature~\cite{Hamaguchi2015}, or under an excess static pressure, can be used to slow down dissolution sufficiently to allow enough time for experiments, or to drive a slow bubble growth for which solvent pockets are no longer an issue~\cite{Ando2019}. Imaging oscillating or collapsing bubbles around $\sim 1$~MHz finally poses the challenge of resolving the radial dynamics, which requires acquisition rates of $\sim10^7$ images per second. Commercial instruments now meet these requirements~\cite{Kuroda2018}, and will become more widely used as their price becomes more affordable.}

A complex issue of linear bubble oscillations is the nature of the quantities $G$ and $\eta$ obtained from fitting the resonance curve. Even though the two techniques are conceptually equivalent, resonance curves obtained from frequency sweeps obviously contain information at multiple frequencies, whereas resonance curves obtain from radius sweeps are single-frequency measurements. It is therefore tempting to identify the material properties $G$ and $\eta$ fitted from radius sweep data with the storage modulus $G'(\omega)$ and loss modulus $G''(\omega)$ measured from classical oscillatory rheology at the acoustic angular frequency $\omega$. In this case, the effective frequency of quantities fitted to frequency sweep resonance curves become unclear. However,  Equation~(\ref{eq:resonance_curve}) is derived assuming a Kelvin-Voigt, linear rheology of the surrounding medium. In classical rheology, this implies that the storage modulus $G'(\omega) = G$ and loss modulus $G''(\omega) = \eta \omega$ are prescribed at all frequencies. Fitting the resonance curves then constrains $G'$ and $G''$ at every frequency based on experimental data covering a limited frequency range --frequency sweep resonance curves-- or even covering a single frequency --radius sweeps. This contrasts with classical oscillatory rheology, for which measurements cover three decades in terms of frequency, allowing easy discrimination between different rheological models. Resonance curve data should then ideally be complemented by measurements from another technique to validate (or question) the choice of the rheological model~\cite{Jamburidze2017}, keeping in mind that the rheological models presented here (neo-Hookean, Maxwell, Kelvin-Voigt) are rather crude approximations of the experimental behaviour of real soft materials.

The last limitation is related to the onset of shape oscillations, which break down the spherical symmetry assumption used to derive most of the equations derived previously. Shape oscillations have been observed both under moderate amplitude bubble oscillations and during cavitation events in both Newtonian fluids and visco-elastic solids~\cite{Brujan2006,Hamaguchi2015,Saintmichel2020}. Stability of spherical bubbles during during cavitation events and oscillations of moderate amplitude have been studied respectively by Gaudron~\emph{et al.}~\cite{Gaudron2020} and Murakami~\emph{et al.}~\cite{Murakami2020}. The latter article quantifies the shift in shape mode number due to medium elasticity and the impact of viscosity on the critical oscillation amplitude above which shape oscillations occur. Agreement with the available, very limited experimental data~\cite{Hamaguchi2015} was found to be good. More experimental data would be valuable to validate the model and eventually use shape oscillation data to obtain rheological information from soft materials.

\subsection{Opportunities for high-frequency rheology}

As complex fluids are structured from intermediate ($\sim\mu$m) to molecular ($\sim${\AA}) sizes, they also show a broad distribution of relaxation time scales which can be probed through linear oscillatory rheology. In polymer solutions and wormlike micelles, these time scales are directly related to spatial physical quantities of the polymers such as the chain persistence length and the entanglement or crosslink density. In particulate suspensions, gels and emulsions, the linear oscillatory spectra contain valuable information on the onset of the glass transition or rigidity percolation~\cite{Larson1999}. Commercial rotational rheometers cannot provide high frequency measurements ($\geq 50$~Hz) due to inertial effects of the rheometer head. Time-temperature superposition has then been widely used in polymer solutions and melts to fill this particular gap in experimental data~\cite{Larson1999}; yet, the technique cannot be applied for out-of-equilibrium (arrested or glassy) systems or those undergoing phase transitions. Dedicated piezoelectric and microelectromechanical systems (MEMS) offer direct high-frequency rheological measurements, as recently reviewed by Schroyen~\emph{et al.
}~\cite{Schroyen2020}. Traditional microrheology techniques, relying on the motion of passive or active tracers embedded in the fluid~\cite{Mason2000}, have also been developed to complement relaxation spectra both at very small and large frequencies. Bubble dynamics offer an opportunity to extend the range of available techniques for active microrheology.

Many industrial flows, such as jetting, injection moulding, or lubrication, and natural flows, such as sneezing~\cite{Scharfman2016}, apply strains to complex fluids at high frequency and beyond their linear deformation, that is, at high strain rates. Such transformations break and reorient their microstructure, usually leading to shear-thinning behaviour. The associated, very high strain rates $\dot \epsilon \geq 10^4$~s$^{-1}$ once again exceed the capabilities of rotational rheometers due to sample expulsion, particle migration and edge fracture. Pressure-driven flows, such as capillary rheometers, are a viable alternative to rotational geometries as they do not present any free surface in the fluid region of interest, yet they require careful pressure corrections due to end effects and potential wall slip~\cite{Macosko1994}. Bubble microrheology, in particular based on bubble collapse, is unaffected by wall slip, since the strain is applied by a movement normal to the bubble-fluid interface, and meets both the high-frequency and high-strain criteria to provide insights on high-strain rate rheology of many complex fluids.

\add{Lastly, bubble dynamics, collapse and cavitation can generate shear waves in viscoelastic materials when the bubbles are located close to a boundary~\cite{Montalescot2016,Tinguely2016, Rapet2019}, because the strain profile around the bubble is no longer symmetric. These waves have been observed using echography~\cite{Montalescot2016}, particle tracking~\cite{Tinguely2016} or birefringence~\cite{Rapet2019} techniques. The propagation speed and attenuation of shear waves can provide a direct measurement of the high-frequency material properties $G$ and $\eta$ of the material~\cite{Catheline2004}, which have been compared to low-frequency measurements obtained using a rheometer in~\cite{Tinguely2016}.}

\subsection{Opportunities for extensional deformation of complex fluids}

Extensional deformation is the main mode of deformation during spinning, rolling, convergent or divergent die injection and spraying in industrial processes~\cite{Macosko1994} but also in biological flows, e.g. during sneezing~\cite{Scharfman2016}. Extensional deformation, in contrast with simple shear, strongly imposes the orientation of the fluid elements, leading for instance to the well-known coil-stretch transition and extensional thickening of polymer solutions and the high extensional viscosity of rigid fibre suspensions~\cite{Macosko1994}. 
In yield-stress fluids, the yield stress in extensional flow is known to differ from the one obtained in simple shear~\cite{Niedzwiedz2010,Zhang2018} due to the non-linear normal stress differences arising in the material even below yielding~\cite{deCagny2019,Varchanis2020}. As solid boundaries in a geometry impose shear flows, true extensional flows work with stress-free boundary conditions, for instance by examining the thinning and breakup dynamics of fluid filaments, following either the separation of two end plates~\cite{McKinley2000} or the Rayleigh-Plateau instability in jet flows~\cite{Keshavarz2015}. As the strain in the filament leads to an exponential decrease of their cross-section, these techniques are limited in terms of maximum applied strain. Bubble microrheology is then particularly interesting to locally apply very strong extensional strains, given by $\epsilon = B_{\theta\theta}(R_0) - 1 = (R/R_0)^2 - 1$ at the bubble edge. Such strains easily exceed $10$ for both collapsing~\cite{Estrada2018} and dissolving bubbles, above the values (typically $6-7$) achieved with capillary breakup rheometry.

\begin{figure}[htb]
    \centering
    \includegraphics[width=80mm]{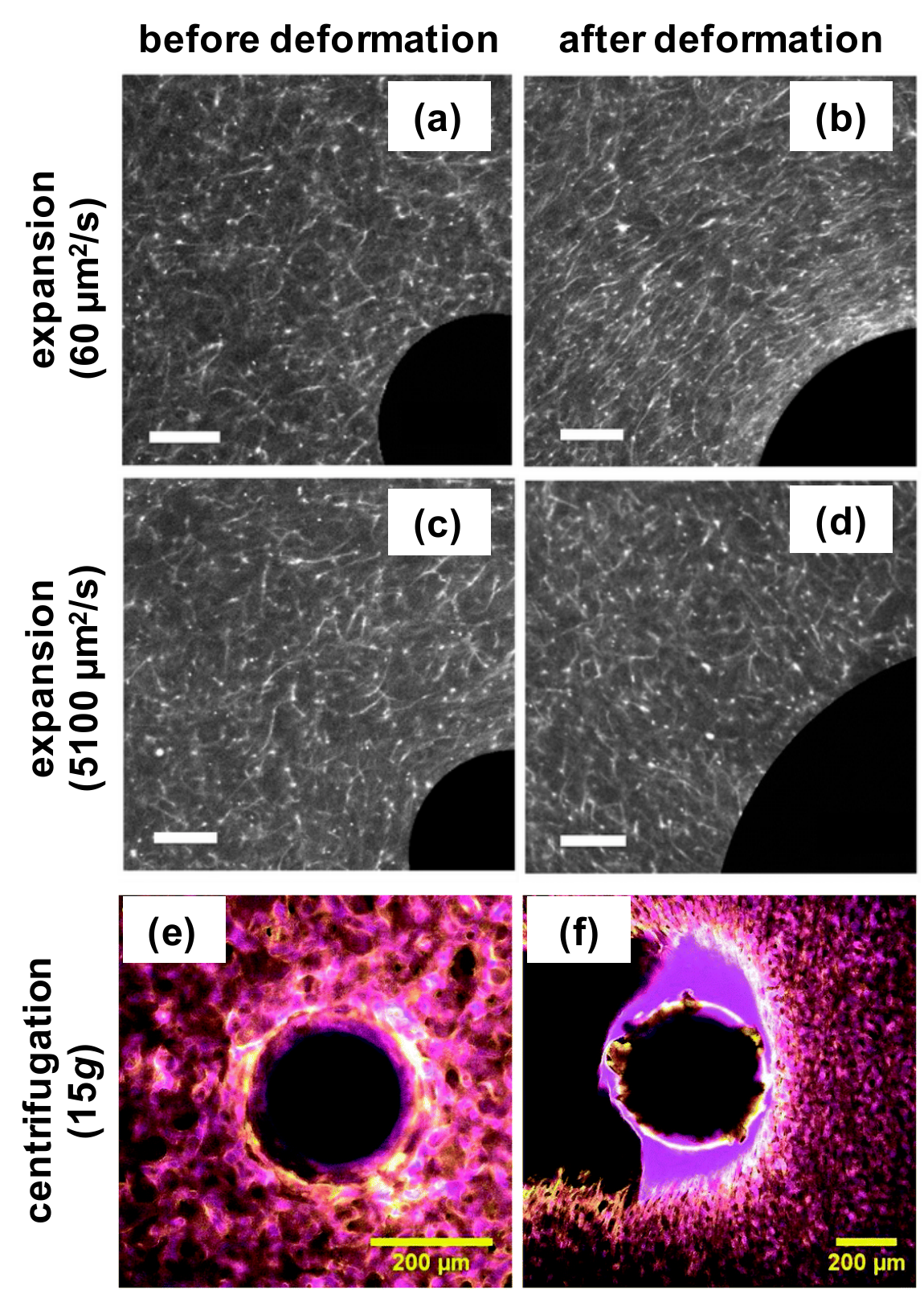}
    \caption{\textbf{Visualisation of bubble-induce deformation of complex fluids.} \textbf{(a-d)} Quasi-static expansion of a bubble in a sparse microcellulose yield-stress fluid. Reproduced with permission from Ref.~\cite{Song2019}. In (a-b), very slow expansion rates allow a relative motion between the solvent and the fibre network and changes in the microstructure are visible close to the expanded bubble. In (c-d), for higher expansion rates, no relative motion is possible and the microstructure around the bubble remains similar before and after inflation. \textbf{(e-f)} Deformation and destruction of a bijel due to bubble displacement during centrifugation. Reproduced with permission from Ref.~\cite{Rumble2016}.\label{fig:microstructure}}
\end{figure}

Mixed, non-uniform flows containing a strong extensional component have been thoroughly investigated in microfluidic cross-slot~\cite{Haward2012,Varchanis2020} and hyperbolic~\cite{Ober2013} geometries. In such geometries, the ratio between simple shear and extension is not known \emph{a priori} and depends on the fluid rheology and the magnitude of wall slip. As a consequence, these geometries are usually complemented with local flow birefringence~\cite{Haward2012} or microparticle image velocimetry~\cite{Varchanis2020} to precisely monitor the time-dependent, non-uniform strain applied to the fluid elements. Direct imaging of the deformation field and evolution of the microstructure is also possible in combination with bubble dynamics experiments. For instance, the reorganisation of a microfibre network around an expanding bubble has been directly visualised for different deformation rates, as shown in \textbf{Figure~\ref{fig:microstructure}a-d}, evidencing a strong reorganisation of the fibre network close to very slowly, quasi-statically expanding bubbles, which progressively disappears as the expansion rate increases. The change in the microstructure of a bicontinuous interfacially jammed emulsion gel (bijel) due to bubble motion during centrifugation has been imaged by confocal microscopy (\textbf{Figure~\ref{fig:microstructure}e-f}). These examples support the notion of simultaneous imaging of the evolution of the microstructure of a complex fluid, in combination with rheological measurements based on the bubble dynamics phenomena described in this paper.

\section{Conclusion}

Spherical bubbles are versatile active probes to characterise the local rheology of soft materials thanks to the purely extensional, localised nature of the deformation field they apply to the material. In this paper we discussed three emerging micro-rheological measurements based on recording the time evolution of the radius of a bubble, during three distinct processes that provide complementary information on the surrounding material. The first process is the slow, quasistatic bubble dissolution dynamics and its potential arrest, which can be used to evaluate the linear elastic modulus $G$ of a material for neo-Hookean or Kelvin-Voigt materials, or to obtain an estimate the critical stress $\sigma_{\rm Y}$ of yield-stress fluids. A second technique exploits linear bubble oscillations driven by an acoustic wave. Examining the bubble response in the time or in the frequency domain provides
an estimate from high-frequency data of both $G$ and the solvent viscosity $\eta$, for an equivalent Kelvin-Voigt material. Lastly, the violent bubble collapse process offers additional, deeper insights on the finite-strain rheology of soft materials at extremely high strain rates (typically $10^6$~s$^{-1}$). Information is obtained by fitting experimental collapse time series to fairly complex, non-linear, finite strain models of bubble dynamics. These modelling steps are needed to estimate the stress field ${\bm \sigma}$ in the material, which is otherwise unavailable, in contrast with classical rheology measurements. Dissolution and collapse processes impose high extensional strains close to the bubble boundary and may be a powerful technique to examine extreme phenomena such as the extensional thickening of polymer solutions due to individual chain uncoiling; in particular since the direct bubble imaging can easily be combined with particle image velocimetry techniques or flow birefringence.
We also discussed outstanding challenges of bubble-based microrheology, for instance controlling the stress history and the process of bubble injection. More work is also needed to combine more realistic models to describe soft materials with the governing equations of bubble dynamics, especially under high-frequency, high-strain rate deformation. Under such conditions, the use of simple models, for instance neo-Hookean Kelvin-Voigt solids, should at least be better justified given the numerous deviations from such a behaviour observed in classical rheology~\cite{Storm2005}.

\section*{Conflict of Interest Statement}
The Authors declare no conflicts.

\section*{Acknowledgements}
This work was supported by a European Research Council Starting Grant [grant number 639221] (V.G.).

\bibliographystyle{apsrev}
\bibliography{./biblio}

\end{document}